\documentclass[final,1p,times]{elsarticle}
\usepackage{graphicx}
\usepackage{graphics}
\usepackage{amssymb}

\begin{document}

\begin{frontmatter}

\title{A curious spacetime entirely free of centrifugal acceleration}

\author[nkd1,nkd2]{Naresh Dadhich}
\ead{nkd@iucaa.ernet.in}



\address[nkd1]{Centre for Theoretical Physics, Jamia Millia Islamia,
New Delhi 110025, India}

\address[nkd2]{Inter-University Centre for
Astronomy \& Astrophysics, Post Bag 4, Pune 411 007, India}


\begin{abstract}
In the Einstein gravity, besides the usual gravitational and
centrifugal potential there is an additional attractive term that
couples these two together. It is fun to enquire whether the latter
could fully counteract the centrifugal repulsion everywhere making
the spacetime completely free of the centrifugal acceleration. We
present here such a curious spacetime metric and it produces a
global monopole like stresses going as $~1/r^2$ in an AdS spacetime.
\end{abstract}

\end{frontmatter}

\noindent PACS numbers :04.20.-q

\medskip

In the Newtonian gravity it is impossible to nullify centrifugal acceleration altogether everywhere. But strange things happen in Einstein's
gravity because of the $3$-space being curved \cite{d1}. As is well known for the gravitational field of a point mass that it is indeed possible
for a particle of non-zero angular momentum to reach the centre $r=0$ which is forbidden in the Newtonian gravity. This happens because there is
an additional  attractive potential arising out of coupling of the gravitational and centrifugal potentials, $2\Phi l^2/r^2$ in the usual
notation. This coupling is facilitated by the space curvature. It is an interesting question to pose whether this additional purely Einsteinian
effect fully counteract the centrifugal acceleration everywhere? This is indeed
so as we show in the following.\\

Let us consider the static spherically symmetric metric,
\begin{equation}
 ds^2 = Adt^2 - \frac{1}{A}dr^2 - r^2d\Omega^2, \,\, A = 1+2\Phi(r).
\end{equation}
This form of the metric follows from the requirement that for radial motion $d^2r/ds^2 = - \Phi^{\prime}$ for timelike particle and $d^2r/d\lambda^2 = 0$ for photon \cite{d1}. Then writing the conserved quantities, $A\dot t = E, r^2\dot\phi = l$, we write
\begin{equation}
 \dot r^2 = E^2 - (1+2\Phi)(l^2/r^2 + \mu)
\end{equation}
where $\mu=1,0$ for timelike and null particle respectively. Differentiating this equation we obtain
\begin{equation}
 \ddot r = -\mu \Phi^{\prime} + \frac{l^2}{r^3}(1+2\Phi-r\Phi^{\prime}).
\end{equation}
Clearly it would be free of centrifugal acceleration everywhere for
\begin{equation}
 1 + 2\Phi - r\Phi^{\prime} = 0
\end{equation}
which readily integrates out to
\begin{equation}
 A = 1 + 2\Phi = kr^2.
\end{equation}

Thus we have the spacetime entirely free of centrifugal force given by
\begin{equation}
 ds^2 = kr^2 dt^2 - dr^2/kr^2 - r^2d\Omega^2
\end{equation}
in which particles can have non-zero angular velocity but no centrifugal radial repultion. This has happened because the effect of coupling of
angular momentum with the gravitational potential perfectly balances the centrifugal acceleration everywhere. This balance occurring at a
particular $r=3M$ and then overpowering it is a familiar aspect in the Schwarzschild field, and that is why there cannot occur circular orbit below $3M$.
The curious thing here is that it is so everywhere and hence complete freedom from centrifugal force.\\

It is obvious that spacetime is not asymptotically flat and particle motion lies in the range $0\leq r^2 \leq E^2/k - l^2$, and hence $E^2/k -
l^2 > 0$ always. The next question arises what kind of stresses this metric gives rise to, or alternatively what is its  physical source? It
turns out that it is indeed the solution of the Einstein equation,
\begin{equation}
 G_{ab} = - T_{ab} - \Lambda g_{ab}
\end{equation}
with $\Lambda = -3k < 0, T^0_0 = T^1_1 = 1/r^2 = \rho = -p_r, p_\theta = T^2_2 = T^3_3 = 0$, where we have set $8\pi G/c^2 = 1$. The global monopole
stresses asymptotically have the same structure falling off as $1/r^2$, and $\rho+p_r = 0$ and the transverse stresses vanishing \cite{bar-vil}. The
global monopole in AdS spacetime would have $A = 1 -\eta^2 +kr^2$ where $\eta$ is the monopole charge. Our solution results from it when $\eta^2 = 1$, the monopole charge being unity. It is therefore a global monopole of unit charge in an AdS spacetime. However it is remarkable that setting charge to unity makes the spacetime completely free of the centrifugal acceleration \\

Also note that at $r=0$, curvatures diverge as $1/r^2$ which means the singularity is weak as integral of $\rho$ over the volume will go as $r$
which would however remain finite at $r=0$. It also marks the horizon of spacetime which is now singular. On the other hand for $k=0$, the curvatures
remain finite and regular but the metric is ill behaved. Normally in such situations the metric could be made regular by a proper choice of coordinates.
But here the cause of trouble is not a coordinate but vanishing of a constant, $k$ which cannot be repaired by any transformation. Thus $k=0$ is not
admissible. This means a global monopole can have $\eta^2=1$ only with $k\neq0$ in AdS spacetime and not otherwise. The most surprising thing is how the
monopole having unit charge is tied to the spacetime being completely free of centrifugal acceleration. \\

The principal difference between Newton and Einstein as argued elsewhere \cite{d1} is the self interaction of gravity for the latter, and it is
brought about by the space curvature while the gravitational force continues to follow the familiar inverse square law. This is also demanded by
light's interaction with gravity because photon can feel gravity only through the curvature of space \cite{d2}. Here again it is the space
curvature which is responsible for this bizzare property. If space was not curved, we would have had $\dot r^2 = E^2/A -l^2/r^2 - \mu$. Then
there won't have been any cancellation of the centrifugal force. It is the curving of space (with $g_{00}g_{11} = -1$, which is demanded by
photon not experiencing radial acceleration \cite{d1}) that gives rise to the possibility of entirely anulling out the centrifugal acceleration. \\

There are many interesting spacetimes in literature with unusual features like the  Kantowski-Sachs cosmological metric \cite{kant} and the
Nariai and Bertotti-Robinson metrics \cite{d3}. The latter two are the product of two spacetimes of constant curvature. The Nariai is an
$\Lambda$ vacuum solution when the two curvatures are equal and it is conformally non-flat while the Bertotti-Robinson is a constant electric
field solution when the two curvatures are equal andopposite and is conformally flat. This is in contrast to the usual $\Lambda$ vacuum
solutions that are conformally flat and the electrovac that are conformally non-flat. The universality of these solutions  for the pure
Lovelock gravity of any Lovelock order $N$ in $d=2N+2$ dimension has also been established \cite{d4}. \\

We have now added one more spacetime of unusual behaviour, however it remains to be seen whether it is simply an amusing curiousity or something
more. Particularly the association of the centrifugal free character with the global monopole charge being unity, $\eta^2 = 1$, is it a mere
coincidence or does it reflect something deeper and profound? \\

\bibliographystyle{elsarticle-num}

\begin{thebibliography}{90}
\bibitem{d1} N. Dadhich, Einstein is Newton with space curved, arxiv:12060635.
\bibitem{bar-vil} M. Barriola and A. Vilenkin, Phys. Rev. Lett. {\bf 63}, 341 (1989).
\bibitem{d2} N. Dadhich,Subtle is the gravity, gr-qc/0102009, Pramana {\bf 77} 1 (2011), Int. J. Mod.Phys. {\bf D20}, 2739 (2011).
\bibitem{kant} R. Kantowski and R. K. Sachs, J. Math. Phys. {\bf 7}, 443 (1966).
\bibitem{d3} N. Dadhich, On product spacetime with 2-sphere of constant curvature, gr-qc/0003026.
\bibitem{d4} N. Dadhich and J. M. Pons, On universality of the pure Lovelock gravity for the generalized Nariai and Bertotti-Robinson solutions, under
preparation.

\end{thebibliography}

\end{document}